\documentclass[acmlarge,nonacm]{acmart}
\AtBeginDocument{\settopmatter{printacmref=true}}

\AtBeginDocument{%
  }

\usepackage{graphicx}
\usepackage{subcaption}
\usepackage{lipsum}
\usepackage{tcolorbox}
\usepackage[table,dvipsnames]{xcolor}
\usepackage{enumitem}
\usepackage{pifont}
\usepackage{makecell}
\usepackage[table]{xcolor}
\usepackage{xcolor,colortbl}

\newcommand{\name}{CVWeap}

\begin{document}

\title{The Weaponization of Computer Vision: Tracing Military-Surveillance Ties through Conference Sponsorship}

\author{Noa Garcia}
\affiliation{%
  \institution{The University of Osaka}
  \country{Japan}
}

\author{Amelia Katirai}
\affiliation{%
  \institution{University of Tsukuba}
  \country{Japan}
}

\settopmatter{printacmref=false} 
\renewcommand\footnotetextcopyrightpermission[1]{} 

\begin{abstract}
Computer vision, a core domain of artificial intelligence (AI), is the field that enables the computational analysis, understanding, and generation of visual data. Despite being historically rooted in military funding and increasingly deployed in warfare, the field tends to position itself as a neutral, purely technical endeavor, failing to engage in discussions about its dual-use applications. Yet it has been reported that computer vision systems are being systematically weaponized to assist in technologies that inflict harm, such as surveillance or warfare. Expanding on these concerns, we study the extent to which computer vision research is being used in the military and surveillance domains. We do so by collecting a dataset of tech companies with financial ties to the field’s central research exchange platform:  conferences. Conference sponsorship, we argue, not only serves as strong evidence of a company’s investment in the field but also provides a privileged position for shaping its trajectory. By investigating sponsors’ activities, we reveal that $44\%$ of them have a direct connection with military or surveillance applications. We extend our analysis through two case studies in which we discuss the opportunities and limitations of sponsorship as a means for uncovering technological weaponization.
\end{abstract}

\maketitle

\section{Introduction}
The history of science and technology is closely tied to the history of war, with military interests long contributing to key technological innovations such as semiconductors, electronic computers, and the Internet~\citep{mowery2010military}.
Artificial intelligence (AI) is no exception: initially developed through DARPA\footnote{Defense Advanced Research Projects Agency, a research agency of the United States Department of Defense.}-funded initiatives~\citep{dobson2023birth}, the field has since received large investments from military agencies, notably  the United States (US) Department of Defense\footnote{Recently renamed to Department of War~\cite{olay2025trump}.} (DoD)~\citep{widder2024basic}. While, recently, private companies have increasingly become intermediaries in this relationship \cite{richardson2022military,katz2020artificial},  there have been attempts to obscure their involvement, as evidenced by the 2018 leak of internal Google emails detailing deliberate efforts in hiding the company's role in Project Maven~\citep{fang2018leaked,crawford2021atlas}, a Pentagon program for automatic object detection in drone footage \citep{dod2017_projectmaven}. 

In practice, this secrecy coincided with a period where the technology’s immaturity kept military AI on the margins of both application and mainstream ethical discourse within the field. The poor performance of AI on fundamental benchmarks such as MNIST for digit recognition~\citep{lecun2002gradient} or PASCAL VOC for object detection~\citep{everingham2010pascal}, prevented, for decades,  reliable deployment in commercial settings, let alone warfare. As a result, the potential harms of military AI remained largely theoretical. This has changed with the advancement of the technology. With large  models achieving strong performance across several tasks~\citep{brown2020language,radford2021learning,liu2023visual,li2023}, multiple reports reveal that AI is now being integrated into the battlefield. To take just two documented examples: in Ukraine, AI-powered autonomous drones are being deployed for lethal operations~\citep{boffey2025killing}, while in Gaza, the Israeli military has  used AI-driven systems to automatically generate assassination target lists~\citep{abraham2024lavender}.

Although the military applications of AI extend to most of its domains, recent scholarly discussions predominantly centre around autonomous weapons systems (AWS) in general or the role of large language models (LLM) in military planning and decision-making in particular \citep{sinmons2024position,kaffee2023thorny,rivera2024escalation}. Here, we focus the analysis on a critical domain for  weaponization: computer vision, the discipline dedicated to interpreting visual data through pattern recognition and machine learning. While computer vision has potential beneficial applications like medical image analysis for early cancer detection \citep{litjens2017survey} or biodiversity conservation \cite{reynolds2025potential}, the technology can be easily weaponized to contribute to systems that inflict harm, exert control, or facilitate repression. 
Specific examples include object detection for autonomous lethal drones \cite{fang2018leaked}, target identification in aerial imagery \cite{chen2024robust}, and facial recognition systems for mass surveillance \cite{amnesty2023israel}. As recently showed by \citet{kalluri2025computer}, computer vision research disproportionately focuses on humans and human activities, directly enabling surveillance infrastructures, which, for example, can become tools for operationalizing and normalizing institutional violence in border regimes \cite{alma99149780320802021}. Despite the well-documented harms compiled by \citet{gebru2024beyond}, the field still lacks both the critical analysis and methodological tools to quantify its role in harm infliction.

In this work, we explore whether the weaponization of computer vision is systematically embedded in the field, challenging the concept of scientific neutrality and the framing of harmful applications as incidental rather than structural.
Through empirical investigation, 
we trace sponsorship at major conferences and examine the miliary and surveillance connections of leading computer vision institutions and corporations. 
We collect a dataset from public sources, namely the \name{} dataset, following a four-part methodology:

\begin{enumerate}[leftmargin=*]
    \item We compile a list of $469$ companies and research institutes that sponsored major computer vision conferences between 2004 and 2024. Since conferences serve as the field's primary hubs for knowledge exchange, we argue that sponsorships offer privileged positions for influencing research directions through economic support, hiring, and showcasing events. 
    \item We enrich the collected list with metadata for each sponsor, including its country of origin, founding year, industry, and corporate relationships.
    \item We develop an annotation framework for categorizing corporation's involvement in the weaponization of computer vision, which classifies whether a sponsor is related to military or surveillance applications.
    \item We assign each sponsor an involvement profile based on their public engagement with military and surveillance uses, ranging from organizations where these activities constitute their core operations to conglomerates  with multiple activities.
\end{enumerate}

The \name{} dataset is used to address the following research questions (RQ):

\begin{enumerate}[label=RQ\arabic*.]
    \item How much are computer vision conference sponsors involved in the military and surveillance domains?
    \item How has the presence of sponsors related to military and surveillance evolved over time?
    \item Which countries dominate the computer vision sponsorship landscape and which countries have more sponsors involved in the military and surveillance domains?
    \item Which computer vision subfields have the highest propensity for the military and surveillance domains?
\end{enumerate}

Our findings show that $44\%$ of sponsors in the \name{} dataset have documented ties to the military and surveillance domains. Among these, about $30\%$ do not publicly disclose such connections. We complement this quantitative perspective with two case studies that illustrate the role of sponsorship analysis in exposing technological weaponization. In the first one, we show how even research framed as socially beneficial can be weaponized in surveillance contexts, with sponsorship serving as an indicator of such risk. Conversely, in the second case study, we discuss dual-use by exploring how initiatives for humanitarian demining, funded by military organizations, can meet communities needs while risking being repurposed for military operations.

The goal of this paper is to expose the weaponization of computer vision research by quantifying the relationship of companies in the field and the military and surveillance domains. This relationship, which involves nearly half of the companies analysed, may not be obvious to the public: while a small proportion are overtly military and/or surveillance focused, for the rest, this involvement can remain unnoticed and buried within other commercial activities. Ultimately, we aim to alert researchers who may be unaware that their work, however well-intentioned, could be weaponized to enable surveillance, targeting, and death. 
We emphasize, however, that awareness alone is insufficient. Current regulatory approaches, such as the EU AI Act \citep{EU_AI_Act_2024}, exempt military applications from oversight. This creates a  legislative void: when applications of computer vision like emotion recognition are banned for civilian use but permitted for military purposes, the comprehensiveness of the regulation is reduced, and its effectiveness is limited. 
\section{Weaponization in computer vision}
\label{sec:context}
We consider weaponization from the perspective of two key areas: military and surveillance applications of computer vision. 
Below, we provide a brief overview of the links between AI research and the military, clarify our approach to the concept of surveillance, and address the increasing overlap between the two areas. 

\subsection{Military applications}
The relationships between academic research and the military are well-documented \cite{troath2023political}. Within this, the area of AI is no exception. As noted above, the relationship between academia and military interests was responsible for much of the early advancement of the field of AI broadly, through DARPA's support for the field of cybernetics, which was later reconceived of as AI \cite{richardson2022military,crawford2021atlas}. Recent years have seen a shifted balance, such that private interests have overtaken military interests as key drivers of AI research. Yet, at the same time, recent investigation suggests that private entities are increasingly benefiting from lucrative military investments, while developing technologies for military use \cite{richardson2022military}. These activities are rooted in the profitability of what Suchman et al. \cite{suchman2017tracking} refer to as the ``security-industrial complex'', which creates incentives for the development of securitizing technologies such as military and surveillance technologies at both the micro level of individual entities and macro level of governments. At the macro level, the development of AI technologies for military use are seen to be closely interlinked with national priorities and as ``essential requirements in achieving military advantage'' \cite{troath2023political}. In fact, scholars have argued for the ``irresponsibility''  of not making use of AI for military purposes \cite{meerveld2023irresponsibility}. This perspective draws on a popular conception of AI as neutral or objective \cite{meerveld2023irresponsibility}, disembodied or ethereal \cite{crawford2021atlas}. Thus  the use of AI serves to ``make war virtuous'' \cite{richardson2022military}, through its association with values such as precision and rationality, and present a sanitized version of military activity which is at a remove from the human individuals both implementing and targeted by these systems. It is against this background, then, that there are growing incentives for the development of computer vision technologies for military applications.

\subsection{Surveillance applications}
The next consideration is surveillance, and there are competing perspectives on how best to define the concept itself. The term carries more nuance than a more neutral alternative such as monitoring. Our choice of surveillance here is deliberate, and stems from Fuchs' \cite{fuchs2010can} argument that we must distinguish between a neutral and a negative definition of surveillance, where the negative definition sees surveillance as inherently harmful and problematic, as in the tradition of Foucault. This is in contrast to the neutral approach taken by Ball et al. \cite{ball2012routledge}, for example, who define surveillance as ``regard or attendance to a person or to factors presumed to be associated with a person''. It is also noteworthy that they distinguish between strategic and non-strategic surveillance, with strategic surveillance as involving  ``a conscious strategy - often in an adversarial and inquisitorial context to gather information''. By contrast, Fuchs proposes a negative view which sees surveillance as ``a specific kind of information gathering, storage, processing, assessment, and use that involves potential or actual harm, coercion, violence, asymmetric power relations, control, manipulation, domination, disciplinary power''. We align with Fuchs' perspective, as the forms of surveillance under consideration in this paper extend beyond mere observation and are deeply interlinked with issues of power and violence. 

\subsection{Surveillance-military linkages}
Our focus aligns with other conceptualizations of surveillance including Suchman et al.'s \cite{suchman2017tracking} ``technologies of tracking'', described as ``sociotechnical systems in which military and policing operations converge''. This is a key point for this paper: as noted by the authors, while the military and policing were originally separate spheres, the implementation of new technologies and the accelerated development of computer vision technologies for surveillance has led the two areas to increasingly overlap, each expanding in a cycle of mutual reinforcement against the backdrop of a perceived ``unending'' or ``everywhere'' war (Gregory, 2011, cited in \cite{suchman2017tracking,zehfuss2018war}). This has led to the dual focus of this paper on both military and surveillance uses of computer vision, both of which bring people within the scope of the ``martial gaze'' \cite{bousquet2018eye}. In this context, then, computer vision is an essential technology for the military application of AI, as it supports sensing, imaging, and mapping processes increasingly indispensable for emerging military applications \cite{bousquet2024becoming}.  

\begin{figure*}
    \centering
    \includegraphics[width=0.9\linewidth]{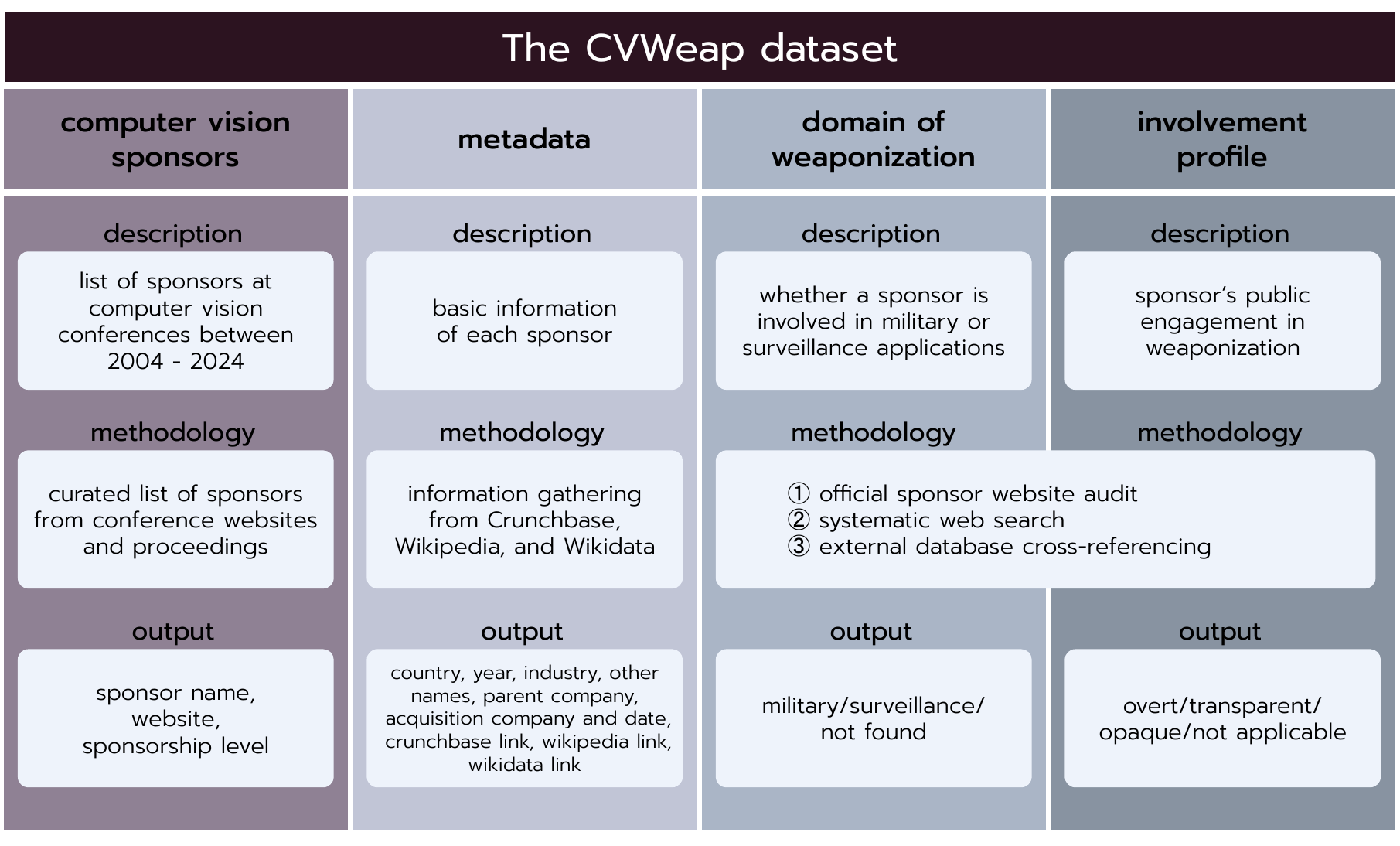}
    \caption{The \name{} dataset collection process consists of four parts: computer vision sponsors, metadata, domain of weaponization, and involvement profile.}
    \label{fig:dataset}
\end{figure*}

\section{Tracing computer vision weaponization}
\label{sec:dataset}
A major challenge in analysing AI weaponization is the lack of transparency regarding research, development, and deployment of military and surveillance applications. Direct methods, such as tracing technologies to specific end products, are often inviable due to the lack of public data, except in cases of whistleblowing \cite{fang2018leaked}. Alternative approaches, thus, need to rely on public sources, such as patents \cite{kalluri2025computer} or military contracts \cite{widder2024basic}. We offer an unexplored via for public data investigation: conference sponsorship. Sponsorship, as we discuss, offers a consistent (yet partial) window into these otherwise opaque relationships.

To investigate weaponization via sponsorship, we collect the \name{} dataset. An overview is presented in Figure \ref{fig:dataset}. The collection process is divided into four parts: computer vision sponsors, metadata, domain of weaponization, and involvement profile, with each of these parts detailed in the following subsections. The dataset is compiled from publicly-available online sources over a ten-month period from November 2024 to August 2025. During this process, some links became inaccessible. 
Despite our best efforts, the dataset also contains gaps where online resources were unavailable from the start of the data collection process, reflecting the fragility of web-based information. As long as accessibility depends on website owners' maintenance, efforts to ensure long-term transparency remain vulnerable.

\begin{table*}[t]
\footnotesize
\setlength{\tabcolsep}{8pt}
\centering
\caption{Details on computer vision conferences and sponsorship in the \name{} dataset.}
\begin{tabular}{@{}l r l r r r r r}
\toprule
& & & editions & \textbf{editions} & sponsors & sponsors & \textbf{sponsors} \\
conference & since & location & 2004-2024 &  \textbf{\name{}} & entries & unique &  \textbf{\name{}} \\
\midrule 
CVPR &  1983 & United States & $21$ & $\textbf{17}$ & $929$ & $342$ & $\textbf{334}$\\ 
ICCV &  1987  & International & $10$ & $\textbf{9}$ & $228$ & $162$ & $\textbf{162}$\\ 
ECCV &  1990  & Europe & $11$ & $\textbf{8}$ & $245$ & $178$ & $\textbf{149}$\\ 
\midrule 
overall &  & & $42$ & $\textbf{36}$ & $1,402$ & $503$ & $\textbf{469}$ \\
\bottomrule
\end{tabular}
\label{tab:conferences}
\end{table*}

\subsection{Computer vision sponsors}
\label{sec:sponsors}

The \name{} dataset compiles tech companies and research institutes engaged in computer vision research and their relationship with military and surveillance applications. To systematically collect companies and institutes engaged in the field, we use conference sponsorship as a proxy. Computer vision conferences are the primary platform for knowledge exchange within the field, a role that has expanded  from small university seminars in the 1980s to mass events with over 12,000 attendees in 2024 \cite{cvpr2024}. The involvement of industry has experienced a similar growth, extending beyond mere logo depiction to active participation through paper submissions, product showcases, and recruiting. A company's decision to sponsor such an event is, therefore, a signal of its commitment to and investment in the computer vision community. Tracing conference sponsorships allows us to collect a list of companies and research institutes that, while not exhaustive, is part of the corporate computer vision ecosystem.

We collect sponsors from the three premier computer vision conferences:\footnote{As recognized by academic consensus and supported by the 2025  Google Scholar ranking in Computer Vision \& Pattern Recognition \citep{scholar2025}.} 
the IEEE/CVF Conference on Computer Vision and Pattern Recognition (CVPR) held annually since 1983; 
the IEEE/CVF International Conference on Computer Vision (ICCV), held biennially in odd-numbered years since 1987; 
and the European Conference on Computer Vision (ECCV), held biennially in even-numbered years since 1990. The historical location of the conferences reflects the field's geographical centres of influence, determining who can most easily attend and benefit from in-person knowledge exchange: 
CVPR is held almost exclusively in the United States, except for the 2023 edition, which was in Vancouver, Canada; ICCV rotates between America, Europe, and Asia; and ECCV is defined as an European conference held in Europe, with the notable exception of the 2022 edition that was in Tel Aviv, Israel. Due to the COVID-19 pandemic, all  three conferences were virtual in 2020 and 2021.

For each conference edition, we manually collect the list of sponsors from the conference's official website or proceedings. We gather data for all the online-available editions, missing CVPR 2004 -- 2007, ICCV 2009, and ECCV 2006, 2014, and 2016. For each sponsor, we collect its name, website, and sponsorship level.\footnote{Conferences typically offer tiered sponsorships that provide benefits according to their cost, such as exhibitor space size. Although not all sponsorship rates are publicly available, they appear to have increased over time. For example, sponsorship tiers at ECCV 2004 ranged from €$3,000$ to €$10,000$, while by ECCV 2020, they had increased to £$5,000$ to £$17,500$.} 
In total, we collect $1,402$ sponsors from $36$ editions, with  duplicates between editions. We manually unify sponsor names and apply automatic text matching for deduplication, obtaining $503$ unique sponsors. We further filter down the list by excluding sponsors unrelated to the use of computer vision, such as academic journals or tourism organizations. Specifically, we remove sponsors falling into the following six categories: publishers and academic journals (e.g., Elsevier), universities (e.g., University of Cambridge), public administration bodies (e.g., Seoul Metropolitan Government), non-tech companies (e.g., Louis Vuitton), media organizations (e.g., What's AI Podcast), and professional institutions (e.g., Institution of Engineering and Technology). The excluded categories account only for $6.8\%$ of the unique sponsors, with the final dataset containing $469$ tech companies and research institutes. Statistics per conference are provided in Table \ref{tab:conferences}.

\subsection{Metadata}
Apart from the name, website, and sponsorship level, we broaden the information about each sponsor with additional metadata. Metadata is manually collected from three sources: Crunchbase,\footnote{\url{https://www.crunchbase.com/} [Last accessed: August 2025].} a privately-owned database of businesses; Wikipedia,\footnote{\url{https://www.wikipedia.org/} [Last accessed: August 2025].} an online crowdsourcing encyclopedia; and Wikidata,\footnote{\url{https://www.wikidata.org/} [Last accessed: August 2025].} a crowdsourcing knowledge base. The collected metadata consists of 1) basic information about the sponsor: country of origin, year of foundation, industry category, and other names of the company; 2) relationship to other companies: parent company (if any), and acquisition company and date (if acquired); and 3) links to the sources endpoints: crunchbase, wikipedia, and wikidata. 
This metadata structure facilitates diverse analytical approaches, such as geographical via country of origin and sectorial via industry category, a label provided by Crunchbase to classify companies. Additionally, company relationships enable analysing corporate networks.

\begin{table*}
\footnotesize
\centering
\caption{External databases used to cross-reference sponsors. \textit{Size} indicates number of companies in the database, and \textit{refs} whether the database contains source references.}
\begin{tabular}{@{}l p{8cm} r r c}
\toprule
name  & scope & size & domain & refs \\
\midrule 

\makecell[tl]{AI War Cloud Database \citep{aiwarcloud}} & AI weapons and companies developing them. & 40 & military &  \ding{51} \\[5pt]

\makecell[tl]{AIGS Index \citep{feldstein2019global}} & AI and big data surveillance use for 179 countries and companies involved.  & 99 & surveillance & \ding{51}\\ [5pt]

\makecell[tl]{Atlas of Surveillance \citep{atlas}} & Surveillance technologies deployed by law enforcement across the US. & 342 & surveillance & \ding{51} \\ [5pt]

\makecell[tl]{Investigate \citep{investigate}} & Companies involved in Israeli military occupation. & 255 & military & \ding{51} \\ [5pt]

\makecell[tl]{SIPRI Arms Industry Database \citep{sipri}} & Arms-producing and military services companies. & 256 & military & - \\ [5pt]

\makecell[tl]{Surveillance Watch \citep{surveillancewatch}} & Companies that develop technologies to enable tracking individuals.   & 684 & surveillance & \ding{51}\ \\

\bottomrule
\end{tabular}
\label{tab:databases}
\end{table*}

\subsection{Domain of weaponization}
\label{sec:domain}
We annotate whether each sponsor is involved in military or surveillance applications according to documented evidence.
Sponsors are assigned one of the three mutually exclusive labels: \textit{military}, indicating involvement in applications directed toward warfare, defence, or armed conflict; \textit{surveillance}, indicating involvement in applications directed toward monitoring, data collection, and observation of populations or individuals; or \textit{not found}, indicating that no documented evidence in the previous two domains is found. For sponsors involved in both military and surveillance applications, the military label takes precedence, as surveillance in these cases is often a subset of military operations.

\begin{figure*}[t]
\vspace{7pt}
\centering
\includegraphics[width=0.98\linewidth]{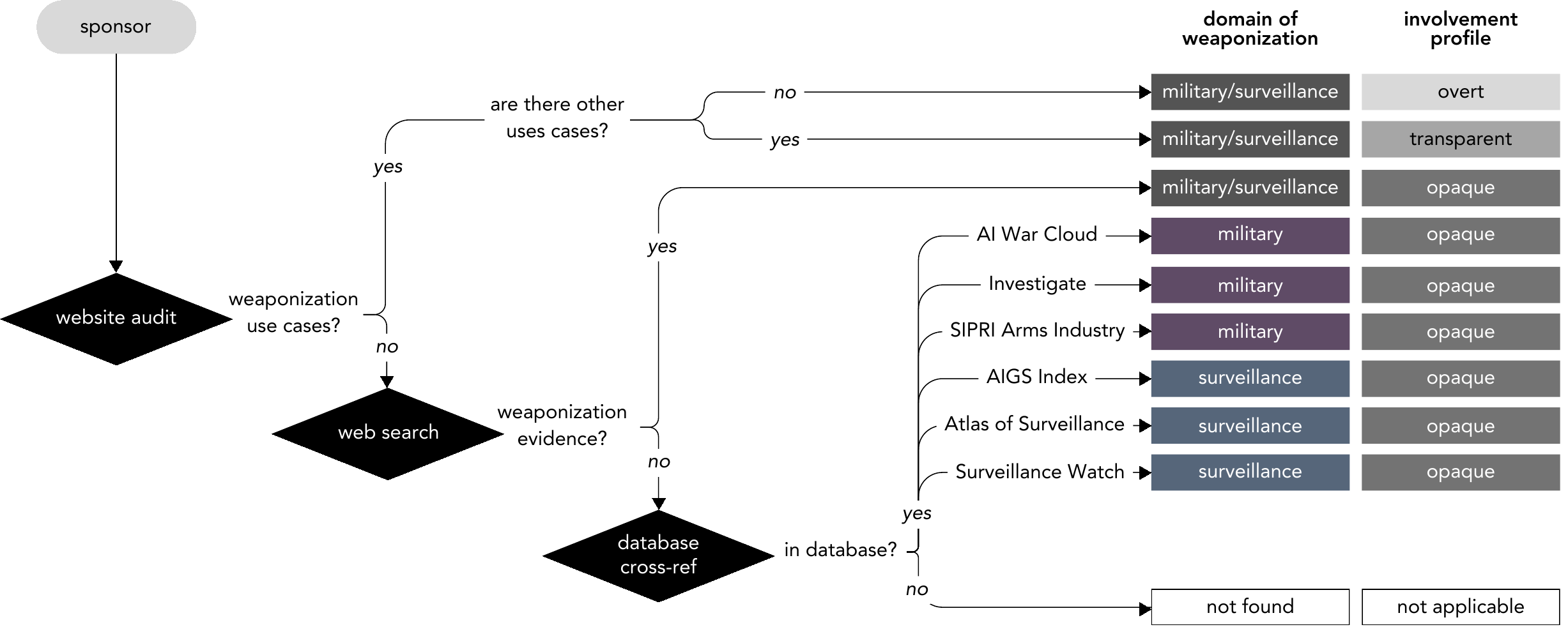}
    \caption{Domain of application and involvement profile annotation flowchart.}
    \label{fig:flowchart}
\end{figure*}

Auditing a sponsor's involvement in weaponization presents numerous challenges, as evidence may be non-public or obscured. Acknowledging the inherent limitations of public sources, our investigation follows a three-step methodology consisting of (1) official website audit, (2) systematic web search, and (3) external database cross-referencing. Each of these steps are detailed below:

\begin{enumerate}[leftmargin=*]

\item{\textbf{Official website audit.}} First, we examine the sponsor's official website for advertised use cases related to military or surveillance applications. For military applications, common terminology includes \textit{defence} or \textit{national security}, often accompanied by imagery evocative of military operations. For surveillance, companies may use the term \textit{surveillance} directly. While advertised military use cases are typically easy to identify, surveillance is sometimes obscured under alternative terminology such as \textit{smart cities}. 

\item{\textbf{Systematic web search.}} For sponsors not previously labelled as \textit{military} or \textit{surveillance}, we conduct systematic web searches using the queries ``$\langle$sponsor name$\rangle$ military'', ``$\langle$sponsor name$\rangle$ defence'', and ``$\langle$sponsor name$\rangle$ surveillance''. We inspect the top retrieved results and search for solid documented evidence, such as journalistic reports or data from funding agencies.

\item{\textbf{External database cross-referencing.}} As information hosted on websites is inherently fragile and it may be removed specially when companies are acquired or terminated, we conduct an additional verification step by cross-referencing sponsors against external, trustworthy databases. We use six publicly available databases, detailed in Table \ref{tab:databases}, that document companies involvement in military and surveillance applications: three related to military (AI War Cloud Database \citep{aiwarcloud}, Investigate \citep{investigate}, and SIPRI Arms Industry Database \citep{sipri}), and three related to surveillance (AIGS Index \citep{feldstein2019global}, Atlas of Surveillance \citep{atlas}, Surveillance Watch \citep{surveillancewatch}). The selection of databases criteria includes the scope of the database, transparency on the reasoning for inclusion, and the availability of references. Each database provides a complementary perspective: some focus on arms manufacturers, while others on surveillance; some are US-centric, while others offer a global overview. The AI War Cloud Database and the AIGS Index specially focus on AI. For cross-referencing, we automatically match sponsor names in the \name{} dataset against the names of the companies in the six external databases with TFIDF \cite{salton1988term}. We then manually verify that the matches belong to the same company and remove false positives. If a sponsor is found in a database, its domain of weaponization is annotated with the domain in Table \ref{tab:databases}. 
\end{enumerate}

The annotation process is primarily conducted by a single author. To ensure consistency in the application of the surveillance definition discussed in Section \ref{sec:context}, sponsors labelled as \textit{surveillance} are subjected to a second review by the other author.

\subsection{Involvement profile}
\label{sec:profile}
As there are different degrees in which a company can be involved in the weaponization of computer vision, we additionally annotate each sponsor with an involvement profile. This profile is used to indicate a company's public posture regarding its involvement in military or surveillance applications, and it consists of four mutually exclusive classes: \textit{overt}, indicating that the company's primary products are for military or surveillance applications; \textit{transparent}, indicating that the company publicly advertises military or surveillance use cases alongside other products; \textit{opaque}, indicating that military or surveillance involvement is documented through external sources but not publicly advertised by the company; or \textit{not applicable}, where no documented evidence of involvement is found. The involvement profile label is assigned directly based on the source of the weaponization evidence. Evidence obtained via official website audits leads to an \textit{overt} or \textit{transparent} profile, depending on whether military or surveillance is their sole focus or one of several advertised use cases, respectively. Evidence found only through systematic web searches or external databases cross-referencing results in an \textit{opaque} profile. 
A summary of the annotation process is provided in Figure \ref{fig:flowchart}.

\section{Analysing sponsorship weaponization}
\label{sec:analysis}

We analyse the data collected in the \name{} dataset by formulating four research questions about  computer vision conference sponsors and their relationship with military and surveillance applications: 1) the degree  of computer vision sponsors involvement in the military and surveillance domains, 2) the sponsorship evolution over time, 3) the regional prevalence of the sponsors, and 4) the differences across computer vision subfields.

\subsection{RQ1: How much are computer vision sponsors involved in military and surveillance?}

\begin{figure*}[h]
\vspace{7pt}
\centering
\includegraphics[width=0.9\linewidth]{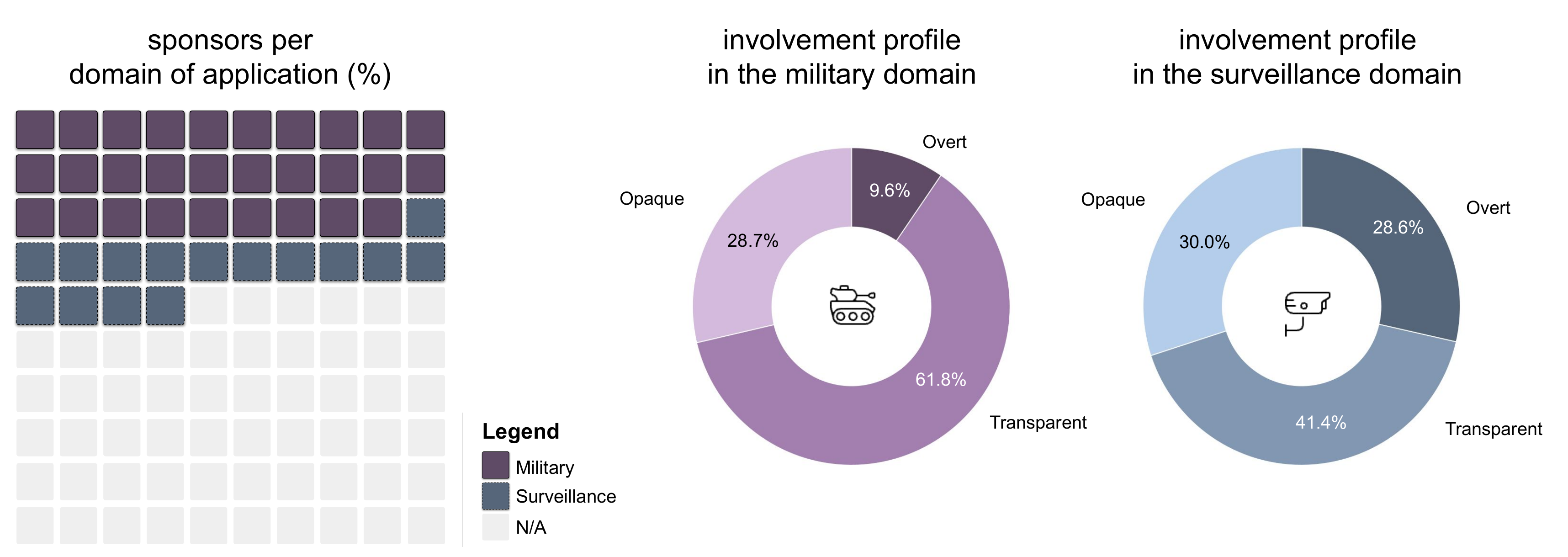}
    \caption{Domain of application and involvement profile statistics in the \name{} dataset. Left: percentage of sponsors per domain of application, with each square representing $1\%$ of the sponsors. Middle: involvement profile (overt, transparent, or opaque) of the sponsors in the military domain. Right: involvement profile of the sponsors in the surveillance domain.}
    \label{fig:involvement}
\end{figure*}

Statistics for the domain of application and the involvement profile in the \name{} dataset are presented in Figure \ref{fig:involvement}. The waffle chart in Figure \ref{fig:involvement} (left), where each square represents $1\%$ of the dataset, shows that $44\%$ of the sponsors  have been found to be involved in the military or surveillance domain. This prevalence, however, may contrast with public perception. As the pie charts in Figure \ref{fig:involvement} (middle and right) show, only a small proportion of the involved sponsors have an overt involvement profile, meaning their main business is military and/or surveillance. The majority of the sponsors have either a transparent (information is public but not necessarily widely known) or an opaque (connections are not disclosed) profile, which can obscure the extent of the field's ties to these industries.

In particular, of the $469$ sponsors in the dataset, $136$ ($29\%$) are involved in the military domain and $70$ ($15\%$) in the surveillance domain. Their involvement profiles slightly differ. 
While there are nearly twice as many sponsors involved in military than surveillance, surveillance-related sponsors are more likely to make it a central part of their business (overt profile). In contrast, both domains show a consistent proportion for the opaque profile: approximately $1$ out of $3$ sponsors hide their military or surveillance uses. With an overt profile, computer vision conferences have been sponsored, among others, by Anduril, AnyVision, Axon, BAE Systems, Lockheed Martin, Rafael, and Raytheon, as well as received financial support directly from the US National Security Agency (NSA).

\subsection{RQ2: How has the presence of sponsors related to military and surveillance evolved over time?}

\begin{figure*}[h]
    \centering
    \begin{subfigure}{0.6\textwidth}
        \centering
        \includegraphics[width=\textwidth]{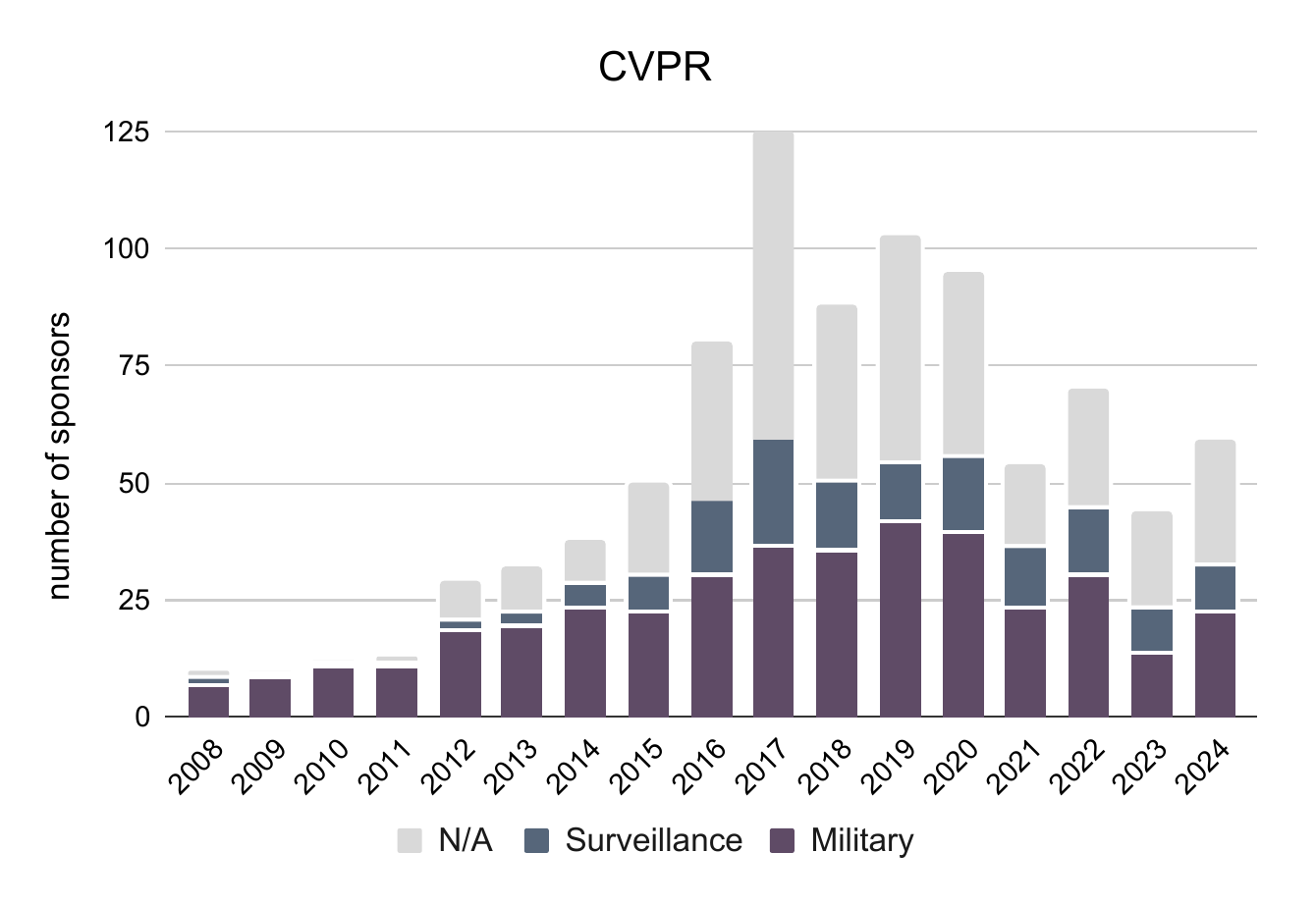}
    \end{subfigure}
    \hspace{-10pt}
    \begin{subfigure}{0.33\textwidth}
        \centering
        \vspace{0pt}
        \begin{subfigure}{\textwidth}
            \centering
            \includegraphics[width=0.9\textwidth,height=0.18\textheight]{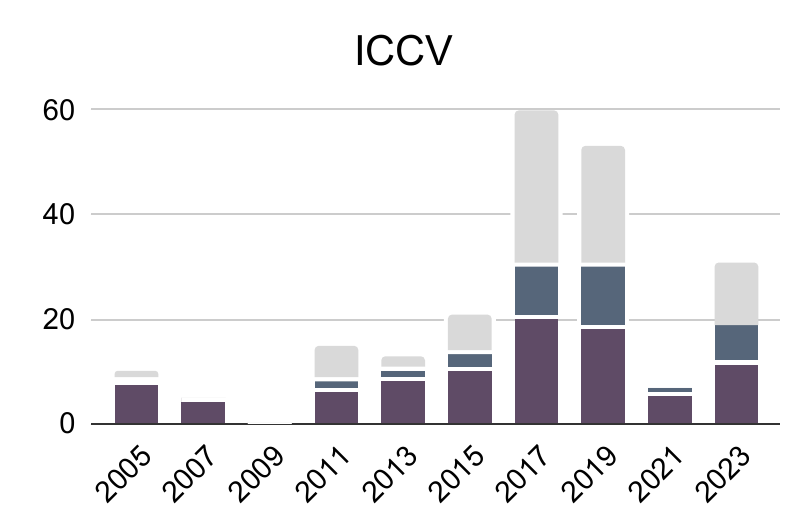}
        \end{subfigure}
        
        \vspace{0.3cm}
        
        \begin{subfigure}{\textwidth}
            \centering
            \includegraphics[width=0.9\textwidth,height=0.18\textheight]{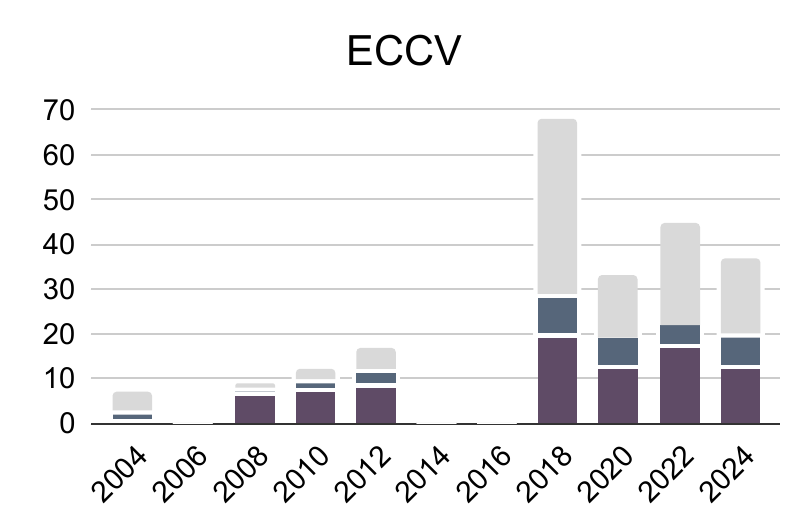}
        \end{subfigure}
    \end{subfigure}
    \caption{Number of sponsors per edition in each of the three main computer vision conferences.}
    \label{fig:sponsors_year}
\end{figure*}

Figure \ref{fig:sponsors_year} shows the number of sponsors for the $20$-year period of collected conference editions, color-coded by their domain of application: military, surveillance, or not found (N/A). There is a big growth in the total number of sponsors from the earliest available data until a peak in 2017, after which a declining trend begins. Despite missing editions (CVPR 2004--2007; ICCV 2009; ECCV 2008, 2014, 2016), the trend is consistent in all three conferences. 

Analysis of the domain of application over time reveals existing patterns. Sponsors involved in military applications have been a constant presence in computer vision conferences since the earliest available data. For example, in CVPR 2009--2011, over $78\%$ of the small pool of sponsors has military ties. As the total number of sponsors grows post-2011, the count of military-involved sponsors increases in absolute terms, but is outpaced by the number of sponsors with no known ties to either military or surveillance. In the surveillance domain, the trajectory is different. While there is almost no presence of surveillance-related sponsors in the early years, beginning around 2014, their presence exhibits an upward trend. This growth aligns with the technological advancements in deep learning that improved face recognition accuracy \cite{taigman2014deepface,schroff2015facenet}.

\subsection{RQ3: Which countries dominate the computer vision sponsorship landscape and which countries have more sponsors involved in the military and surveillance domains?}

\begin{figure*}[h]
\vspace{7pt}
\centering
\includegraphics[width=0.65\linewidth]{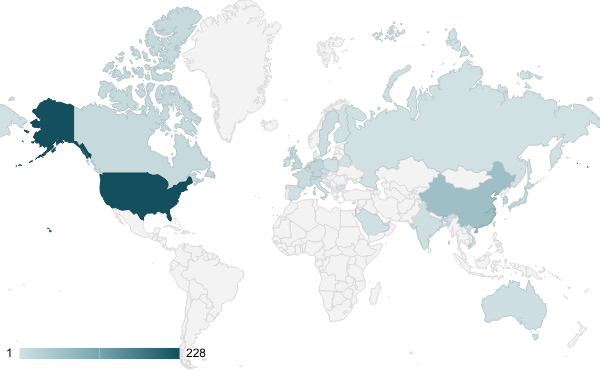}
    \caption{Where are computer vision sponsors from? Number of sponsors per country in the \name{} dataset.}
    \label{fig:map}
\end{figure*}

The map in Figure \ref{fig:map} illustrates the global distribution of sponsors in the \name{} dataset. The geographic concentration of sponsors largely aligns with the traditional boundaries of the Global North, with the additions of China, India, Saudi Arabia, and the United Arab Emirates. Almost half of the sponsors ($48.61\%$) were founded in the US, which is consistent with the large number of conference editions hosted there ($14$ out of $36$). China ranks as the second-largest contributor, with its total number of sponsors accounting for  $28\%$ of those based in the US.

\begin{figure*}[h]
\centering
    \includegraphics[width=\linewidth]{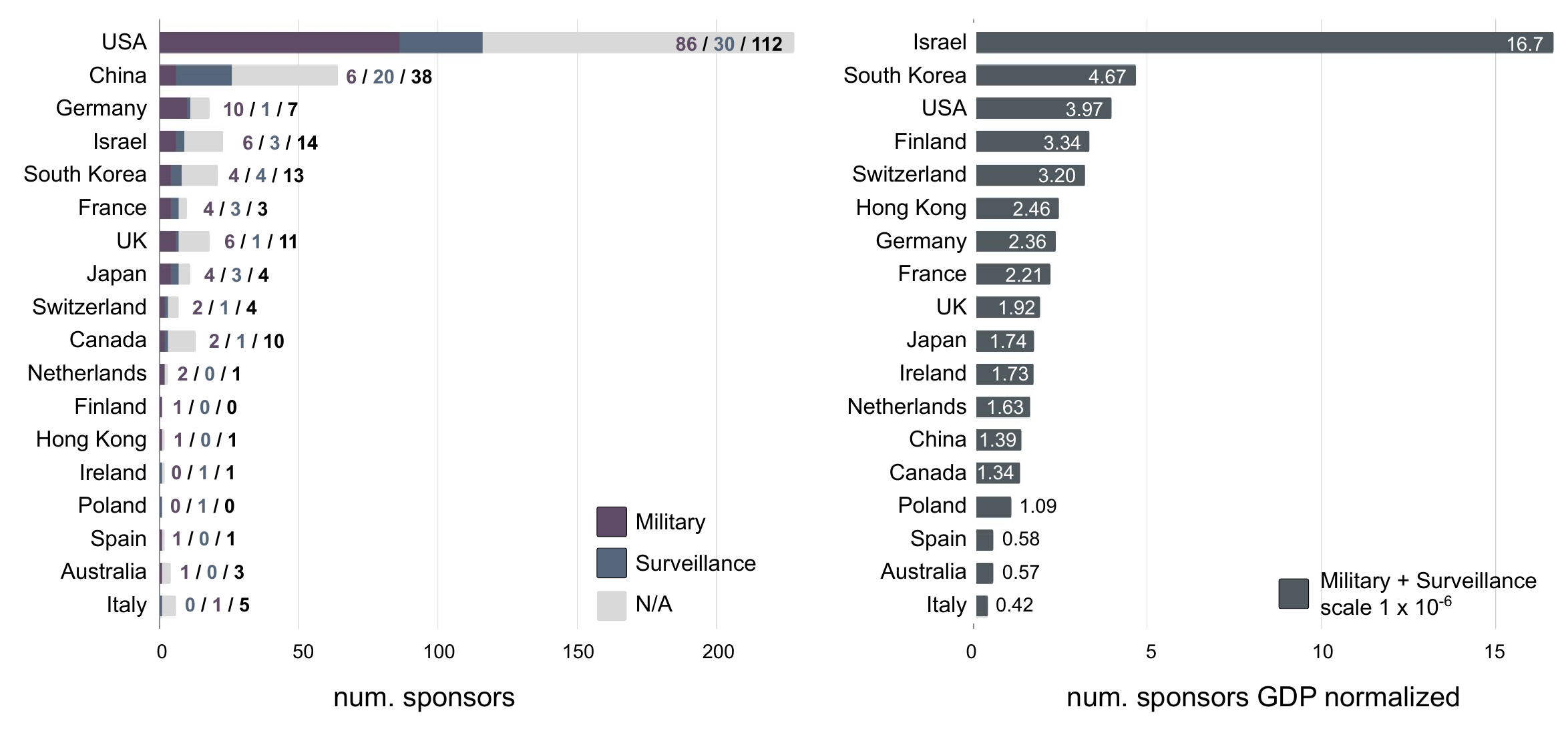}
    \caption{Number of sponsors per country with at least a sponsor involved in the military or surveillance domain. Left: number of sponsors per country and domain of application, sorted in descending order by the number of military + surveillance sponsors. Right: number of sponsors in the military + surveillance domain normalized by the country's GDP.}
    \label{fig:countries}
\end{figure*}

Statistics stratified per domain of weaponization are presented in Figure \ref{fig:countries} (left). The United States dominates in absolute terms:  it has more sponsors involved in the military domain than the total from any other single country, and it also accounts for the largest number of sponsors in the surveillance domain, followed closely by China. When considering the relative distribution of sponsors within each country, China, France, and Japan have the highest proportion of their sponsors active in surveillance, while Germany and France have the highest proportion involved in military applications. 
A direct comparison between the two largest contributors, the US and China, shows that the US has a larger proportion of its sponsors involved in the military domain ($37.7\%$ vs. $9.38\%$), whereas a larger share of Chinese sponsors are involved in surveillance ($13.16\%$ vs $31.25\%$). Overall, the US has a greater absolute number of sponsors involved in both combined domains ($116$ vs. $26$) and a higher percentage ($50.88\%$ vs. $40.63\%$).

While Figure \ref{fig:countries} (left) shows the number of sponsors per country, absolute counts naturally make larger economies dominate. To compare sponsor concentration relative to economic size, in Figure \ref{fig:countries} (right), we normalize the combined military and surveillance sponsor count against each country's 2024 GDP \cite{WorldBankGDP}. The United States and China, which lead in absolute numbers, rank 3rd and 13th, respectively, in the GDP-normalized view. 
The adjusted ranking is dominated by Israel, which places first by a vast margin, with a value $3.5$ times greater than the second-ranked country, South Korea. Israel's outsized lead in this metric confirms its role as a leading developer and exporter of military and surveillance AI. This status is intrinsically linked to what critics and human rights groups have described as a ``testing ground'' for these technologies \cite{loewenstein2024palestine}, extensively documented in the context of the Occupied Palestinian Territory \cite{amnesty2023israel,OHCHR2022Settlements}.

\subsection{RQ4: Which computer vision subfields have the highest propensity for the military and surveillance domains?}

To analyse the relationship between computer vision subfields and the military and surveillance domains, we rely on the industry category labels in the \name{} dataset, which are originally extracted from Crunchbase. While these labels describe broad industry categories, we identify a subset that corresponds to traditional computer vision subfields. We filter out categories with fewer than three sponsors and manually select those relevant to computer vision, resulting in a  set of $14$ categories.\footnote{Facial Recognition, Data Collection and Labeling, Video, Generative AI, Virtual Reality, Computer Vision, Robotics, 3D Technology, Augmented Reality, Machine Learning, Image Recognition, Artificial Intelligence, Visual Search, and Video Editing.}

\begin{figure*}[h]
\vspace{7pt}
\centering
\includegraphics[width=0.7\linewidth]{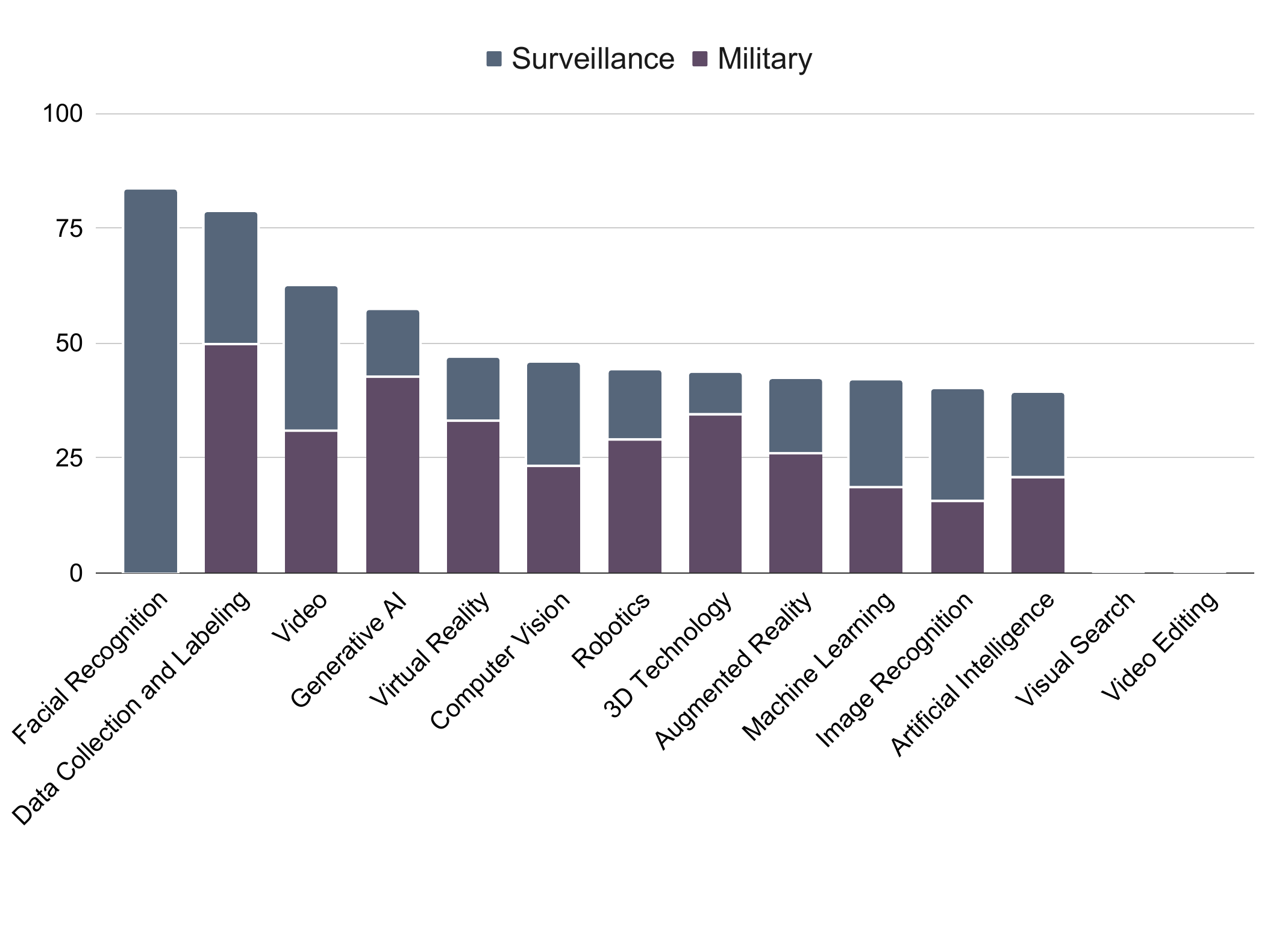}
    \caption{Percentage of sponsors in the surveillance and military domain per computer vision subfield.}
    \label{fig:categories}
\end{figure*}

The results, shown in Figure \ref{fig:categories}, reveal a strong alignment between specific computer vision subfields and weaponization domains. Face Recognition has the strongest connection, with $83.33\%$ of the sponsors in this category involved in surveillance,  confirming the well-documented use of this technology in surveillance systems \cite{andrejevic2022facial,hill2023your,pasquale2020new}. More remarkably, Data Collection and Labelling shows a similarly high level of involvement ($78.57\%$), primarily through contracts with defence and law enforcement agencies, highlighting its role as a foundational enabler for these applications. Furthermore, the subfields of Video and Generative AI also demonstrate significant ties, with more than half of the sponsors in these categories involved in military or surveillance applications. Except for sponsors associated with Visual Search and Video Editing, all categories show a large involvement.

\section{Weaponization beyond quantitative metrics}
With the \name{} dataset, our goal is to quantitatively expose the connection between computer vision sponsors and military-surveillance applications. However, a quantitative analysis alone risks abstracting the human impact of such systems, which have real-world consequences on real people with real lives. Beyond a quantitative analysis, this section discusses two case studies of computer vision uses for militarization and surveillance. In the first one, we expose how ``fair'' face recognition systems can also be weaponized against marginalized communities. In the second one, we analyse demining efforts based on computer vision technologies, acknowledging the complexity of dual-use systems.

\subsection{Case Study 1: Fair face recognition}
In parallel to their main programs, major computer vision conferences host competitively selected workshops on specialized topics. At ECCV 2020, held online, there were 45 accepted workshops, among them the Fair Face Recognition and Analysis workshop. In the workshop's official website,
its motivation was framed as advancing societal good:

\begin{quote}
This workshop will focus on bias analysis and mitigation methodologies, which will result into more fair face recognition and analysis systems. These advances will have a direct impact within society's equality of opportunity. In this proposal we plan to provide a comprehensive up to date review on fair face recognition and analysis research. We find of crucial interest to centralize ideas, discuss them and push the field to advance towards more fair systems for the good of society.\footnote{\url{https://chalearnlap.cvc.uab.es/workshop/37/description/} [Last accessed: December 2025]}
\end{quote}

The workshop was associated with the Fair Face Challenge, a face recognition competition in which different teams tried to obtain the best accuracy, while reducing bias, in a newly collected face verification dataset. The objective of the challenge was for participants to ``develop their fair face verification method aiming for a reduced bias in terms of gender and skin color''.\footnote{\url{https://chalearnlap.cvc.uab.es/challenge/38/description/} [Last accessed: December 2025]} The challenge became relatively popular, with $151$ participants and over $1,800$ submissions. The competition had a cash prize for the best ranked teams, directly sponsored by AnyVision, which offered an award of €$1,000$ for the first place and €$500$ each for the second and third. AnyVision also participated in the creation of the dataset by internally annotating the images, as described in the challenge official report.

AnyVision (later rebranded as Oosto) is an Israeli surveillance company specialized in facial recognition technology, listed in the AIGS Index \citep{feldstein2019global}, the Atlas of Surveillance \citep{atlas}, Surveillance Watch \citep{surveillancewatch}, and our own \name{} dataset. Its primary clients include Israeli authorities, and its systems are deployed across the Occupied Palestinian Territory \citep{loewenstein2024palestine}. In this context, facial recognition functions, as Amnesty International documents \cite{amnesty2023israel}, as tools of ``automated apartheid'', controlling and segregating Palestinian and Arab populations. Since facial recognition is known to be less accurate for non-white individuals \cite{buolamwini2018gender}, technologies aimed at monitoring Arab communities would naturally be interested in developing algorithmic ``fairness'' interventions to function effectively on their intended targets. Through this case study, thus, we confirm two key insights: first, that even applications aimed at fairness can be weaponized, as warned by \citet{benjamin2019race}; and second, that tracing corporate sponsorship may contribute to uncover such weaponization pathways.

\subsection{Case Study 2: Computer vision for demining}
Our second case study is centred in the use of computer vision for humanitarian demining. Landmines pose a severe and lasting threat to war-affected regions, rendering land impassable for people and animals, unusable for agriculture, and artificially dividing communities that may otherwise be geographically close. According to the Landmine Monitor 2025 report \cite{international2025landmine}, in 2024, landmines caused $6,279$ casualties in 52 countries and areas, with $90\%$ of the victims being civilians and nearly half of those being children. Demining is, therefore, a critical humanitarian necessity to restore safety, livelihoods, and social cohesion. However, detecting and removing landmines is dangerous, slow, and complex.

Remote sensing (i.e., techniques that gather information from a distance, typically via satellites, aircrafts, or drones) has the potential to assist in landmine detection while reducing risk for involved humans. By using computer vision techniques, remote sensing images can be processed to detect, classify, and localize buried landmines. For example, \citet{filipi2022honeybee} monitored designated areas with drones to identify clusters of trained honeybees that indicated the presence of explosive residues, a significant technical challenge given the insects' small size and rapid movement. Alternatively, \citet{vivoli2025holomine} proposed the use of holographic imaging and created a synthetic dataset of microwave holographic images to address data scarcity and improve detection and classification of buried landmines.

While landmines are inherently military artifacts, their removal is a civilian and humanitarian need. An inspection of the funding sources of the aforementioned demining efforts reveals sponsorship by the North Atlantic Treaty Organization (NATO) through its Science for Peace and Security (SPS) Programme, which aims to support ``tailor-made, civil security-relevant activities that respond to NATO’s strategic objectives''.\footnote{\url{https://www.nato.int/en/about-us/organization/nato-structure/science-for-peace-and-security-hub/science-for-peace-and-security-programme} [Last accessed: December 2025]} This military sponsorship illustrates the dual-use nature of the technology: the same systems developed to clear mines in a post-conflict zone for humanitarian purposes can be repurposed in an active warzone to clear enemy territory for military advancement. While sponsorship inspection can be useful to uncover technological weaponization, it may also overlook legitimate civilian applications, underscoring the importance of analyses centred in the necessities and agency of affected communities.
\vspace{-5pt}
\section{Limitations of sponsorship analysis} 
As discussed in the preceding case studies, sponsorship analysis offers both opportunities and limitations to investigating technological weaponization. The \name{} dataset was collected to expose the connections between computer vision and its military and surveillance applications, and our quantitative analysis shows an important degree of involvement by computer vision corportations in these domains. However, several limitations must be acknowledged. Importantly, while our data demonstrates correlation, it cannot establish causation between sponsorship and specific weaponization pathways. Furthermore, the dataset is not exhaustive (as no dataset can be) and potential revealing attributes are not included. For example, data about sponsors size in terms of revenue or employee count, which may reveal connections with respect to a company's involvement profile, was not collected. We also acknowledge its Western-centric bias, as it favours English-language documentation and lacks access to sources in other languages. Finally, the \name{} dataset allows for further analyses than the ones presented in this paper, such as sponsorship level and extended corporate networks.
\vspace{-5pt}

\section{Conclusion}

We investigated the weaponization of computer vision research within military and surveillance domains. We did so by collecting the \name{} dataset, a compilation of companies and research institutes sponsoring the three major conferences in the field between 2004 and 2024. Alongside basic organizational information, we annotated each sponsor based on its publicly documented connections to and promotion of military and surveillance applications and funding. Our quantitative analysis showed strong ties: $44\%$ of sponsors in the dataset are involved in military or surveillance activities, and approximately $30\%$ of those do not actively disclose these connections to the public. We complemented the quantitative analysis with two cases studies, illustrating how sponsorship analysis can potentially contribute to expose technological weaponization while also revealing the inherent complexity of dual-use research.

\section*{Acknowledgements}
This work was financially supported by JSPS KAKENHI No.~22K12091 and No.~23H00497. We thank Leonardo Impett for his support during the conceptualization of this reasearch and Piera Riccio, Giorgos Tolias, Vladan Stojnić, and Evangelos Kazakos for their valuable comments on the manuscript.

\section*{Author Contributions}
Noa Garcia (Conceptualization, Data Curation, Methodology, Validation, Formal analysis, Visualization, Writing - Original Draft, Writing - Review \& Editing) and Amelia Katirai (Investigation, Validation, Writing - Original Draft, Writing - Review \& Editing).

\section*{Generative AI usage statement}
Generative AI played no role in the intellectual and methodological foundations of this work, including its conception, literature review, contextualization, data collection, data annotation, data analysis, and visualization. An LLM was used exclusively by the first author during manuscript preparation for post-writing proofreading.

\bibliographystyle{ACM-Reference-Format}
\bibliography{sample-base}

\end{document}